\documentstyle[aps,prb,multicol,epsf]{revtex}
\begin{document}  
\draft
\title{Nonequilibrium and Parity Effects in the Tunneling Conductance 
of Ultrasmall Superconducting Grains}                      
         
\author{Oded Agam and Igor L. Aleiner}   
\address{NEC Research Institute, 4 Independence Way, Princeton, NJ
08540}
\maketitle
\begin{abstract}
Recent experiment on the tunneling spectra of ultrasmall
superconducting grains revealed an unusual structure of the lowest
differential conductance peak for grains in the odd charging states.  
We explain this behavior by nonequilibrium ``gapless'' excitations associated
with different energy levels occupied by the unpaired electron.  
These excitations are generated by inelastic cotunneling.  
\end{abstract}     

\begin{multicols}{2}

The electron-electron interaction strongly
affects the tunneling transport through ultrasmall metallic grains. 
Many phenomena of grains in the normal state are
described by the  orthodox model in which the interaction takes a
simple form $E_{\mbox{int}} =Q^2/(2C)$, where $Q$ and $C$ are the total
charge and capacitance of the system respectively \cite{review}.
This model is successful in describing the Coulomb blockade, 
which is essentially the quantization of the number of
electrons in the grain, when the gate voltage is tuned away from the
charge degeneracy point. Because of this quantization, the zero-bias
conductance of the system vanishes, while the current $I$ as the function
of the source-drain voltage $V$ shows a threshold behavior. Fine
structure of the current--voltage curve was attributed to 
single-electron levels of the system\cite{AverinKorotkov,Leo}.

However, there is a variety of effects which are not described by
the orthodox model.  They may become especially pronounced when the
system is driven out of equilibrium, thereby exploring many
excited states. For normal metallic grains, nonequilibrium 
steady states were shown to lead to clustering of resonances in the
differential conductance due to fluctuations of the
interaction energy among electrons \cite{nonequilibrium}. The
experimental findings for {\em normal} metallic grains \cite{RBT} 
can be summarized as follows:
\begin{enumerate}
\item
Clusters of resonances of the differential conductance spectra appeared 
at high source-drain voltage for which nonequilibrium  
configurations of the electrons become energetically allowed. 
The first peak, however, did not split into several resonances. 
\item
The width of each resonance cluster was found to be much smaller
than the mean level spacing $d$. It is of order  $d/g$, where   
$g \sim 5$ is the dimensionless conductance of the grain.
\end{enumerate}

Recently, Ralph {\em et al.} \cite{Ralph1} measured the tunneling
resonance spectra of ultrasmall {\em superconducting} grains. The number of
electrons in the system was controlled by a gate voltage. The results
of this experiment show that:
\begin{enumerate}
\item For the ground 
state of the grain with an even number of electrons,  
the first peak of the differential conductance is merely shifted by 
the gate voltage $V_g$. The shape of this peak does not change over
a large  interval of $V_g$. 
Contrarily, if the grain contains an odd number of electrons, 
the height of the first peak rapidly reduces with a change of the gate 
voltage. Moreover, a structure of subresonances develops 
on the low-voltage shoulder of this peak, see Fig.~\ref{fig:1}.
\item
The characteristic energy scale between subresonances of the first peak 
is of the order of mean level spacing $d$.
\end{enumerate}

{\narrowtext
\begin{figure}[h]
\epsfxsize=7.8cm
\hspace*{-0.2cm}
\epsfbox{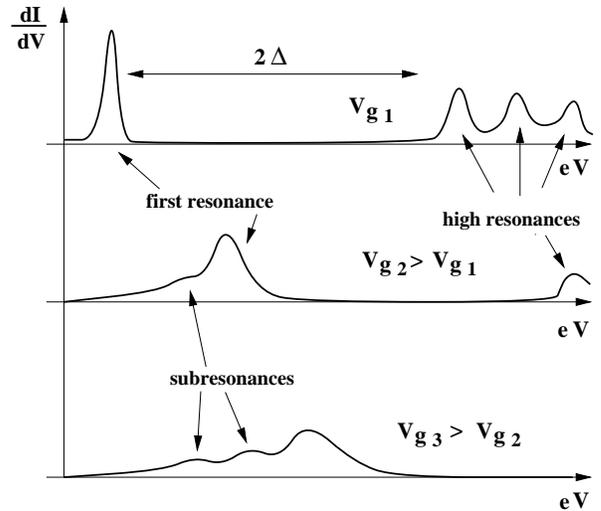} 
\caption{A schematic illustration of the differential conductance of
an ``odd'' superconducting grain as function of the source-drain
voltage $V$, at various gate voltages $V_g$. Higher resonances are
separated by the superconducting gap from the first one, and
subresonances are developed as the first resonance is shifted by the
gate voltage.}
\vspace{-0.2cm} 
\label{fig:1}
\end{figure}
}

These observations contrast the results for the
normal case, nevertheless, it was suggested\cite{Ralph1} that
the substructure of the first peak is still associated with
nonequilibrium steady states of the grain.  The purpose of this paper
is to clarify the origin of these nonequilibrium states and 
the mechanism which generates them. 

The principle difference between odd and even grains is that 
all excitations of the latter are of energy larger than the superconducting
gap $2 \Delta$. Therefore, a source-drain voltage in the range $V < \Delta/e$ 
can not induce nonequilibrium states. Odd grains, on the other hand, 
contain one unpaired electron, which may be shifted to various single 
electron levels with characteristic energy scale smaller than the
mean level spacing $d$. For this reason  even small source -drain voltage
$d<eV<\Delta$ is sufficient to excite the grain.  
The mechanism of excitation is inelastic cotunneling \cite{cotunneling}. 
Tunneling into the excited grain requires less energetic 
electrons, and lead in turn to the substructure on the 
low-voltage shoulder of the of the first resonance, see Fig. 1. 
A closely related problem was considered by Averin and 
Nazarov~\cite{AverinNazarov}, however, their theory assumed 
that relaxation processes prevent the formation of 
nonequilibrium states. For small metallic grains
relaxation processes are very slow\cite{nonequilibrium}, and 
therefore will be neglected in our theory.

To describe the effect quantitatively, 
we construct the master equations governing the time
evolution of probabilities of different electronic configurations of
superconducting grains allowing for second order cotunneling
processes. The solution of these equations for two limiting cases
(one in which two levels participate in the transport, 
and the other when a large number of levels contribute) 
explains the substructure of the first peak of the 
differential conductance illustrated in Fig.~1.

Our model Hamiltonian is given by
$\hat{H}=\hat{H}_0+\hat{H}_{T}+\hat{H}_{\mbox{int}}$.  Here $\hat{H}_0$
describes the noninteracting electrons in the left ($L$) and right ($R$)
leads and in the metallic grain,
\begin{equation}
\hat{H}_0= \sum_{\nu =L,R} \sum_q (\xi_{\nu q}+\mu_{\nu}) 
d^\dagger_{\nu q} d_{\nu q}+ \sum_j \xi_j c^\dagger_j c_j.
\label{H0}
\end{equation}
Single particle energies $\xi_{\nu q}$, are measured relative to the
chemical potentials of the left $\mu_{L}=\alpha eV$ and right
$\mu_R=(\alpha-1)eV$ leads (the numerical coefficient
 $0<\alpha<1$ depends on the capacitances between the grain and the 
leads). Tunneling across the barriers is described by
\begin{equation}
\hat{H}_{T}= 
\sum_{\nu =L,R} \sum_{q,\alpha} T^{(\nu)}_{q\nu}d^\dagger_{\nu q}c_j +
 \mbox{ H.c.},
\label{HT}
\end{equation}
where $T^{(\nu)}_{qj}$ are the tunneling matrix elements. The interaction
Hamiltonian is
\begin{equation}
\hat{H}_{\mbox{int}}=\frac{e^2}{2C} (\hat{N}-N_0)^2
- \lambda \sum_{i,j}  c^\dagger_{i\uparrow} 
c^\dagger_{i\downarrow} c_{j\uparrow} c_{j\downarrow}.
\label{interaction}
\end{equation}
The first term in Eq.~(\ref{interaction}) is the orthodox model for the
repulsive part of the electron-electron interaction:
$C$ is the total capacitance of the grain, 
$\hat{N}=\sum_jc^\dagger_jc_j$ is the number of electrons 
in the dot, and $N_0$ is a continuous parameter controlled 
by the gate voltage.  $N_0$ determines the finite charging energy 
required to insert,  $U_+$, or to  remove, $U_-$, one electron, 
\begin{equation}
U_\pm=\frac{e^2}{C} \left[ \case{1}{2}\pm (N-N_0) \right],\quad
|N-N_0| \leq \case{1}{2}.
\label{U} 
\end{equation}
The second term in Eq.~(\ref{interaction}) drives the grain to the
superconducting state (we assume zero temperature). 

We consider the experimentally relevant case, $e^2/C \gg \Delta$, 
so that the grain has  well defined number of electrons. 
If this number is even
$N=2m$, the ground state energy (we will omit charging part of the energy
and restore it later), $E_{2m}= F_{2m}+2\mu m$, can be calculated in the
mean field approximation\cite{textbook,otherworks} by minimizing 
thermodynamic
potential $F_{2m} = \sum_{k} (\xi_k - \epsilon_k) + \Delta^2/\lambda$
where $\epsilon_k= (\xi_j^2 + \Delta^2)^{1/2}$, with respect to
$\Delta$ ($\partial F / \partial \Delta = 0$), and by fixing the
chemical potential via $2m = -\partial F / \partial \mu$. All the
excited states of even dots are separated from the ground state 
by a large energy, $2\Delta$. 
Considering now the energy spectrum of an odd grain, $N=2m-1$,
we notice that the second term in Eq.~(\ref{interaction}) operates
only within spin singlet states. Therefore, to calculate the
low-lying excited states in this case, we fill the single-electron 
state $j$ with one electron, and then find the ground state of the 
remaining $2m$ electrons with state $j$ excluded from the 
Hilbert space. In the
mean field approximation it corresponds to the minimization of the
thermodynamic potential $F_{2m-1}^{(j)} = \sum_{k\neq j} (\xi_k -
\epsilon_k) + \Delta^2/\lambda + \xi_j$. The excited states with
energies smaller than $\Delta$ are characterized by a single
index, $j$ and will be denoted by $E_{2m-1}^{(j)}$.  
In what follows we will need the energy
cost of introducing an additional electron into the odd state: 
$U_+ +\epsilon_j$, where $\epsilon_j = E_{2m}-E_{2m-1}^{(j)}$. 
Assuming $\Delta \gg d$, one obtains\cite{fluctuations}
\begin{equation}
\varepsilon_j = \mu_{2m} - \frac{3 d}{2} + \frac{\xi_j d}{2 \Delta}
 -\sqrt{ \xi_j^2 + \Delta^2}.
\label{epsilon}
\end{equation}

We turn now to the kinetics of a superconducting grain.
Consider the regime where $U_+ = U \lesssim
\Delta$, $U_- \approx \frac{e^2}{2 C} \gg U$,
and $\frac{e^2}{2 C} \gg \Delta \gg d$. 
We also assume  the conductance of the
tunnel barriers to be much smaller than  $e^2/h$,
and that the source-drain voltage is  small  $eV < \Delta$. 
The simplicity brought to the problem in this regime
of parameters stems from the fact that there is only one available
state with an even number of electrons (because $U_-\gg U_+$ 
one can only add an electron to grain but not subtract one), 
and whenever the grain contains an even number of electrons it is in its  
ground state. This imply that even grains
cannot be driven out of equilibrium state, while for odd grains
tunneling (and cotunneling) takes place via unique state. 

Henceforth, we concentrate on grains with an odd ground state. 
Let us denote by
$P_e$ the probability of finding the grain with an even number of
electrons, and by $P_j$ the probability to find the grain in the odd
state $j$.  Since these states are spin degenerate in the absence of
magnetic field, $P_j$ will denote the sum
$P_{j,\uparrow}+P_{j,\downarrow}$.  The master equations for the
probabilities $P_e, P_j$ have the form
\begin{eqnarray}
&&\frac{ d P_e}{ dt} =  \sum_j \left[ \Gamma_{o \to e}^{(j)} P_{j}
- 2\Gamma_{e \to o}^{(j)} P_e  \right], \label{rate-equations}  \\
&&\frac{d P_j}{dt} \!=\!2\! \sum_{i \neq j}\! \left[ 
\Gamma_{i \to j} P_i \!-\! \Gamma_{j \to i}
    P_j \right]\! +\!2 \Gamma_{e \to o}^{(j)} P_e \!-\! 
   \Gamma_{o \to e}^{(j)} P_j, \nonumber
\end{eqnarray}
where $\Gamma_{o \to e}^{(j)}$ and $\Gamma_{e \to o}^{(j)}$ are the
transitions rates from the odd $j$-th state to the even and from the
even to odd respectively, while $\Gamma_{i \to j}$ is the rate of
transition from the $i$-th to the $j$-th odd states.  Equations
(\ref{rate-equations}) are not linear independent, so they have to
be supplied with the normalization condition $P_e+ \sum_j P_j =1$.
Current in the steady state equals to the electron
flow through, say, the left barrier, and for positive $V$ it is given by
\begin{equation}
I =e \sum_j\left(\Gamma_{o\to e}^{(j)}+\Gamma_{j\to j}
 \right)P_j 
+ 2e\sum_{j\neq i} \Gamma_{j\to
i}^{(j)}P_j.
\label{current}
\end{equation}

\begin{mathletters}
\label{rates}
Transition from
the $j$-th odd state into the even state occurs when $\mu_L >
U+ \varepsilon_j$. The amplitude of this transition is calculated
by first order perturbation theory in the tunneling Hamiltonian
(\ref{HT}). Fermi's golden rule yields
\begin{equation}
\Gamma_{o \to e}^{(j)} = g_L \frac{ u_j^2 \rho_{Lj} d}{2 \pi \hbar}  
\theta(\mu_L-\varepsilon_j-U), 
\label{rate1}
\end{equation} 
where $g_L$ is the dimensionless conductance of the left tunnel
barrier per one spin,
$u_j = (1 + \xi_j/\epsilon_j)/2$ is the coherence factor, $\theta(x)$
is the unit step function, and
$\rho_{Lj}=\Omega |\psi_j(r_L)|^2$, where $\Omega$ is the volume of the 
grain and $\psi_j(r_L)$ is the value of $j$-th single particle wave 
function at the left point contact $r_L$. Energies $\varepsilon_i$ are
given by Eq.~(\ref{epsilon}) and $U=U_+$ is defined in Eq.~(\ref{U}).
Similarly, the rate of transition from even state to $i$-th odd state,
by tunneling of an electron from the dot to the right lead, is
given by     
\begin{equation} 
\Gamma_{e \to o}^{(i)} = g_R \frac{v_i^2 \rho_{Ri} d}{ 2 \pi \hbar}  
\theta(U + \varepsilon_i-\mu_R), \label{rate2}
\end{equation}
where $g_R$ is the dimensionless conductance of the right tunnel barrier,
$v_i = (1 - \xi_i/\epsilon_i)/2$, and $\rho_{Ri}=\Omega |\psi_i(r_R)|^2$,
where $r_R$ is the position of the right point contact.

A change in the occupation configuration of the odd states occurs
via inelastic cotunneling\cite{cotunneling}. 
This mechanism is a virtual process in which
an electron tunnels into $j$-th available level and another electron
tunnels out from the $i$-th level. Calculating this rate by second order
perturbation theory in the tunneling Hamiltonian, one obtains
\begin{equation}
\Gamma_{j\to i} = \frac{g_L g_R d^2 u_j^2 v_i^2 \rho_{Lj} \rho_{Ri}
(eV-\varepsilon_j+\varepsilon_i) }{ 8 \pi^3 \hbar
(U + \varepsilon_j - \mu_L)(U + \varepsilon_i -\mu_R)}
\label{rate3}   
\end{equation}
for $eV>\varepsilon_j-\varepsilon_i$, $\mu_L < U + \varepsilon_j$,
$\mu_R < U + \varepsilon_i$, and zero otherwise.
$\Gamma_{j\to i}$ diverges in the limits $\mu_L \to U + \varepsilon_j$
and $\mu_R \to U + \varepsilon_i$. It signals that a real
transition takes over the virtual one. The region of
applicability of Eq.~(\ref{rate3}) is, therefore, $ U +
\varepsilon_j -\mu_L > \gamma$ and $ U + \varepsilon_i -\mu_R >
\gamma$ where $\gamma \sim gd/4\pi$ is the width of a single particle
level in the dot due to the coupling to the leads, $g= g_L +
g_R$. However,
the interval of biases where Eq.~(\ref{rate3}) is not valid is narrow,
and to the leading approximation in $\hat{H}_T$ our results will be 
independent of this broadening.
\end{mathletters}

{\narrowtext
\begin{figure}[h]
\epsfxsize=8.7cm
\epsfbox{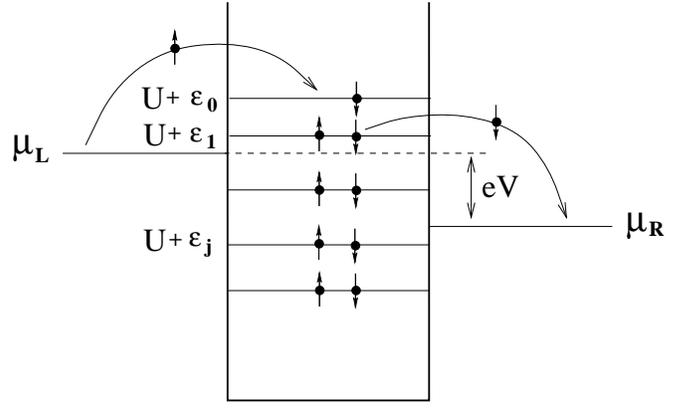} 
\vspace{0.5cm}
 \caption{Inelastic cotunneling process can drive an ``odd'' 
superconducting grain out of its ground state.
In the ground state, the single particle level indicated by 
$U+\varepsilon_0$ is occupied by one electron. 
Excited states are those in which
the unpaired electron is shifted to other single particle levels. 
In a nonequilibrium steady state, low single particle levels become 
available for resonant tunneling, leading to a subresonances structure
of the differential conductance shown in Fig.~1. 
State $j$ shown to be filled with two electrons 
should be understood as a coherent superposition of double occupied and 
empty states with weights $v_j^2$ and $u_j^2$ respectively.
}
\vspace{-0.26cm}
\label{fig:2}
\end{figure}
}

Let us now apply Eqs.~(\ref{rate-equations}) and (\ref{current}) to
describe the appearance of the low-voltage substructure of the first
peak. We will consider two situations: (i) small voltage
such that only one subresonance can emerge on the shoulder of the leading one, 
and (ii) large voltage, $d\ll eV < \Delta$, where 
the substructure of the main resonance consists of 
a large number of subresonances.  

In the first case, the chemical potentials of the
left and right leads are such that transport through the grain
involves only two levels: $\varepsilon_0$ and $\varepsilon_{1}<\varepsilon_0$
corresponding to the ground and the first excited states of the odd
grain. We solve Eqs.~(\ref{rate-equations}) for probabilities $P_0,P_1$
and $P_e$ using Eqs.~(\ref{current}) and (\ref{rates}). There are
two distinct regimes of the source-drain voltage: (1) 
$\mu_L < U + \varepsilon_0$ where transport is dominated by
cotunneling, and (2) $\mu_L \geq U + \varepsilon_0$
where state ``0'' is available for resonant tunneling.
The substructure of the first resonance in the differential conductance
appears in the first regime. Below we show that  
as  $\mu_L$ passes through $U + \varepsilon_{1}$, 
see Fig.~\ref{fig:2}, there is a discontinuity in the current-voltage curve.
In the first regime,  the total current to the leading
approximation in $g_L, g_R$ is a sum
of two contributions, $I\simeq I_{eq} + I_{ne}$.
The first,  $I_{eq}=  e \Gamma_{0\to 0} + 2e \Gamma_{0\to 1}$, is
the equilibrium current coming from  cotunneling. 
The second contribution 
is associated with the nonequilibrium population of state ``1'' 
and is given by
\[
I_{ne} \!=\!2e\Gamma_{0\to 1}   \! \times \!\!\left\{ \begin{array}{ll}
\frac{\Gamma_{1 \to 0}-\Gamma_{0 \to 1} + 2 \Gamma_{1 \to 1}
- 2 \Gamma_{0 \to 0}}{ \Gamma_{0 \to 1}+ \Gamma_{1 \to 0}}
& \mu_L \! < \! U + \varepsilon_1 \\
2\left( 1+ \frac{ \Gamma_{e \to o}^{(1)}} { \Gamma_{e \to o}^{(0)}} \right) 
& \mu_L \!> \! U + \varepsilon_1 \end{array} \right. 
\]
Assuming that the  voltage drop $eV= \mu_L - \mu_R$ is larger than
the energy difference $\tilde{d}= \varepsilon_0 - \varepsilon_1$, 
the jump in the nonequilibrium current is:
\begin{equation}
\delta I_{ne}  =
 c_1 e\frac{ g_L g_R d^2}{ 8 \pi^2 h \tilde{d} }
\left( 1 - \frac{\tilde{d}}{eV} \right)\quad eV \sim 2\tilde{d} 
\label{ne-jump}
\end{equation}
where $c_1= 4 u_0^2 v_1^2 \rho_{L0} \rho_{R1}$
is a constant of order unity. This jump in the nonequilibrium current
leads to the peak in the differential conductance spectra.
Formula (\ref{ne-jump}) has simple interpretation.
Up to numerical prefactors it is a product of two factors: first is the 
probability of finding the grain with an unpaired electron in state 
``1''. It is proportional to $ g_R (d/\tilde{d})
(1-\tilde{d}/eV)$, and increases with
the voltage $V$ and as $\tilde{d}=\varepsilon_{0}-\varepsilon_1
\to 0$. The second factor is associated with the rate in 
which the state ``1'' is filled with an electron, $e g_L d/h$.  

The magnitude of the jump (\ref{ne-jump}) should be compared to the
jump in the current as $\mu_L$ increases above $U+ \epsilon_0$, and
real transition via the even state become allowed.  
To the leading order in $g_L$ and $g_R$, the current
in this regime is 
\begin{equation}
I = c_2 e \frac{ g_R g_L d}{h (g_L + 4 g_R)}, \quad
\mu_L > U + \varepsilon_0,
\label{mainpeak}
\end{equation}
where $c_2$ is a constant of order unity having structure similar to
$c_1$.  Comparing the current jump, $\delta I_{ne}$, with that
associated with the resonant tunneling, $\delta I$, we find
\begin{equation}
\frac{ \delta I_{ne}}{\delta I} \simeq
\frac{ g_L + 4 g_R}{8 \pi^2}, \quad 
eV \sim 2 \tilde{d}.
\end{equation} 
Thus nonequilibrium population of the excited level
of the odd-grain leads to the appearance of a subresonance at small $V$,
however, its height is much smaller than 
that of the main resonance.

We turn now to the second regime of the parameters, $d \ll eV <
\Delta$, in which many levels contribute to the transport. Again,
we focus our attention on the cotunneling regime, $\mu_L <
U+\varepsilon_0$.  We show that the characteristic amplitude
of the subresonances in this regime may become comparable to the 
amplitude of the main peak.
 
To the leading order in $g_L, g_R$, and $d/\Delta$, the steady 
state solution of the rate equations at $\mu_L= U+\varepsilon_1+0$ 
is $P_0 \simeq 1$, while for the other probabilities we have
\begin{equation}
P_e \simeq \frac{\sum_{i \neq 0 }\Gamma_{0\to i}}{\Gamma^{(0)}_{e\to o}},\quad
P_j \simeq 2 \frac{\Gamma_{0 \to j}}{\Gamma_{o \to e}^{(j)}}
+
2 \frac{\Gamma_{e \to o}^{(j)}}{\Gamma_{o\to e}^{(j)}}P_e.
\label{largeN} 
\end{equation}
The characteristic number of states contributing to the current
(\ref{current}) is
large as $\sqrt{\Delta eV}/d$ so that mesoscopic fluctuations of the
tunneling rates and of the inter-level spacings may be neglected. 
Additional large factor, $\sqrt{\Delta eV}/d$, comes from the summation
over the levels in Eq.~(\ref{largeN}), and we find
\begin{equation}
I\simeq \frac{e^2 g_L g_R}{2\pi^2h}\frac{ V \Delta }
{\varepsilon_0-\varepsilon_j}, \quad
\left\{
\begin{array}{l} d \ll eV \ll \Delta \\ \mu_L = U+\varepsilon_j+0
\end{array} \right. .
\end{equation}
Once again, the current jumps  each time $\mu_L$ passes
through  $U +\epsilon_j$.
This jump for large $j$ (but still such that $U + \varepsilon_j 
-\mu_L \ll eV$) scales as $1/j^3$, and the ratio of the jump at $j=1$ 
to the jump at the resonance level (\ref{mainpeak}) is given by
\begin{equation}
\frac{ \delta I_{ne}}{\delta I} \simeq
\frac{ (g_L + 4 g_R)}
{8 \pi^2} \frac{e V \Delta }{\tilde{d} d}, \quad
d \ll eV < \Delta.
\end{equation} 
Noticing that $\tilde{d}=\varepsilon_0-\varepsilon_1 \simeq d^2/ 2\Delta$,
we see that the first subresonance becomes comparable in height 
to the main one
at voltages as small as $eV \approx 4 \pi^2 d^3/ \Delta^2 ( g_L
+ 4 g_R)$.

We conclude by comparing the above results with the experimental data 
of Ref.~[\onlinecite{Ralph1}]. There $\Delta
\approx 5 d$, the conductances in the normal state are $g_R \approx 10
g_L \approx 1/8$, and the leads are also superconducting. The
singularity in the density of states of the leads imply that the
effective conductance is increased by factor of $2-3$. 
Neglecting inelastic processes and the Josephson coupling, these
parameters imply that when $eV \approx 2 d$ the ratio of
the subresonances amplitude to that of the first resonance 
is of order one, while at $eV \sim 2 \tilde{d} \approx d/5$ it is of 
order of 1\%. It implies that first subresonance peak associated with
tunneling into state ``1'' cannot be resolved, both because its amplitude
and its distance from the main peak are too small. However next 
subresonance appear already at distance of order $d$ from the main
resonance, and for $V > 2d$ have an amplitude comparable with the  
main resonance.

We are grateful to B.~L.~Altshuler and L.~I.~Glazman for discussions and
to D.~C.~Ralph for useful communications.

\end{multicols}
\end{document}